\documentclass[journal=nalefd,manuscript=letter]{achemso}
\usepackage[version=3]{mhchem}
\usepackage{graphicx}
\usepackage{bm}

\author{A. M. Burke}
\email{amburke@unsw.edu.au} \affiliation{School of Physics,
University of New South Wales, Sydney NSW 2052, Australia}

\author{O. Klochan}
\affiliation{School of Physics, University of New South Wales,
Sydney NSW 2052, Australia}

\author{I. Farrer}
\affiliation{Cavendish Laboratory, University of Cambridge, CB3 0HE,
U.K.}

\author{D.A. Ritchie}
\affiliation{Cavendish Laboratory, University of Cambridge, CB3 0HE,
U.K.}

\author{A. R. Hamilton}
\affiliation{School of Physics, University of New South Wales,
Sydney NSW 2052, Australia}

\author{A. P. Micolich}
\email{adam.micolich@nanoelectronics.physics.unsw.edu.au}
\affiliation{School of Physics, University of New South Wales,
Sydney NSW 2052, Australia}

\date{\today}

\title {Extreme sensitivity of the spin-splitting and $0.7$ anomaly
to confining potential in one-dimensional nanoelectronic devices}

\begin{document}

\begin{abstract}
Quantum point contacts (QPCs) have shown promise as nanoscale
spin-selective components for spintronic applications and are of
fundamental interest in the study of electron many-body effects such
as the $0.7 \times 2e^{2}/h$ anomaly. We report on the dependence of
the 1D Land\'{e} $g$-factor $g^{*}$ and $0.7$ anomaly on electron
density and confinement in QPCs with two different top-gate
architectures. We obtain $g^{*}$ values up to $2.8$ for the lowest
1D subband, significantly exceeding previous in-plane $g$-factor
values in AlGaAs/GaAs QPCs, and approaching that in InGaAs/InP QPCs.
We show that $g^{*}$ is highly sensitive to confinement potential,
particularly for the lowest 1D subband. This suggests careful
management of the QPC's confinement potential may enable the high
$g^{*}$ desirable for spintronic applications without resorting to
narrow-gap materials such as InAs or InSb. The $0.7$ anomaly and
zero-bias peak are also highly sensitive to confining potential,
explaining the conflicting density dependencies of the $0.7$ anomaly
in the literature.

{\bf Keywords:} one-dimensional system, $g$-factor, quantum point
contact, nanoelectronics.

\end{abstract}
\maketitle

A current focus in nanoelectronics is the development of spintronic
devices where spin is used instead of charge for storage, transfer
and processing of information.~\cite{AwschalomNP07} Non-magnetic
spintronic device elements are highly
desirable;~\cite{AwschalomPhys09} quantum point contacts (QPCs) have
shown great promise being used both as individual spin injectors and
detectors,~\cite{PotokPRL02, FolkSci03, DebrayNN09} and in larger
device structures for studying phenomena such as spin
relaxation~\cite{FrolovPRL09} and ballistic spin
resonance.~\cite{FrolovNat09} The QPC is the quintessential
one-dimensional (1D) electron system, consisting of a narrow
quasi-1D aperture separating two regions of two-dimensional electron
gas (2DEG) in a III-V semiconductor heterostructure. It is typically
defined electrostatically by applying a negative bias to nanoscale
metal gates on the heterostructure surface; its hallmark is a
quantized electrical conductance $G = m G_{0}$, where $G_{0} =
2e^{2}/h$, $e$ is the electron charge, $h$ is Planck's constant, and
$m$ is the number of spin degenerate 1D subbands beneath the Fermi
energy $E_{F}$ of the adjacent 2DEG reservoirs.~\cite{vanWeesPRL88,
WharamJPC88} The spin properties of QPCs are also of fundamental
interest; one example is the conductance anomaly at $G \sim 0.7
G_{0}$,~\cite{ThomasPRL96} where the interplay between 1D
confinement, quasi-bound state formation and exchange-driven spin
polarization are not yet fully understood.~\cite{MicolichJPCM11} The
combined influence of exchange and 1D confinement are also vital to
remarkable behaviors such as spin-charge
separation~\cite{AuslaenderSci05} and the formation of electron
liquid/solid states in 1D electron systems.~\cite{HewPRL08,
DeshpandeNat10}

An important quantity in considering the spin-properties of QPCs is
the effective Land\'{e} $g$-factor $g^{*}$, the constant of
proportionality between the Zeeman splitting of the 1D subbands and
the applied magnetic field. The $g$-factor $g^{*}_{m}$ for each 1D
subband $m$ is easily measured in QPCs.~\cite{PatelPRB91} For spin
injection and detection, it is highly desirable to maximize the
lowest 1D subband $g$-factor $g^{*}_{1}$, which sets the minimum
field required to resolve the spin. The $g$-factor is also a useful
experimental probe of the exchange interaction.~\cite{JanakPR69} The
foundational work on the $0.7 \times 2e^{2}/h$ anomaly showed that
$g^{*}_{m}$ increases from the bulk GaAs value of $0.44$ at $m = 25$
to $\sim 1.15$ at $m < 4$.~\cite{ThomasPRL96} This
`exchange-enhancement' effect,~\cite{ThomasPRB98} also observed in
InGaAs/InP QPCs,~\cite{MartinAPL08, MartinPRB10} is central to the
suggestion that the $0.7$ anomaly is caused by exchange-driven
spontaneous spin-polarization within the QPC.~\cite{ThomasPRL96,
ThomasPRB98} Remarkably, after 15 years of study of the $0.7$
anomaly, little more is known about the dependence of $g^{*}_{m}$ on
QPC confinement potential or the electron density $n$ of the 2DEG in
which the QPC is formed~\cite{MicolichJPCM11}.

Here we study how $g^{*}_{m}$ evolves with $n$ for three samples
featuring two different gate architectures for enacting changes in
density. We pay particular attention to $g^{*}_{1}$ given its
importance for spintronics and the $0.7$ anomaly. We obtain
$g^{*}_{1}$ values as high as $2.8$. This exceeds previous reports
for the in-plane $g$-factor in GaAs QPCs~\cite{PatelPRB91,
ThomasPRL96, CronenwettPRL02, KoopJSNM07, FrolovPRL09}, and
approaches both the perpendicular $g$-factor recently demonstrated
in GaAs QPCs~\cite{RosslerNJP11} and the lower-bound in-plane
$g^{*}$ for InGaAs QPCs~\cite{SchapersAPL07, MartinAPL08,
MartinPRB10}. The link between $g^{*}_{m}$ and density is not
direct; for example, we see {\it opposite} trends in $g^{*}_{1}$
with $n$ for the two architectures. We find that the $g$-factor
$g^{*}_{m}$, and $g^{*}_{1}$ in particular, is sensitive to the
top-gate configuration. This has important consequences for
spintronic applications of QPCs; if high $g^{*}_{1}$ can be obtained
in GaAs QPCs by careful management of the QPC's electrostatic
potential, it lessens the need to use narrow band-gap materials
(e.g., InGaAs, InSb) for which device fabrication is more difficult.
The second key result of our work arises from comparing the density
dependence of the $0.7$ anomaly in the three samples with that of
$g^{*}_{1}$ and the lowest 1D subband spacing $\Delta E_{1,2}$, a
measure of the strength of the 1D confinement. The behavior of the
$0.7$ anomaly in our devices is consistent with the density
dependent spin-gap model~\cite{ReillyPRL02, ReillyPRB05} if the
spin-splitting rate is assumed directly proportional to $\Delta
E_{1,2}$. This highlights the important role that confinement
potential plays in the $0.7$ anomaly, and provides an explanation
for the conflicting reports regarding the density dependence of the
$0.7$ anomaly in earlier literature.~\cite{ThomasPRB98, ThomasPRB00,
NuttinckJJAP00, PyshkinPRB00, HashimotoJJAP01, ReillyPRB01,
WirtzPRB02, ReillyPRL02, ReillyPRB05, LeeJAP06}

\begin{figure}
\includegraphics[width=10cm]{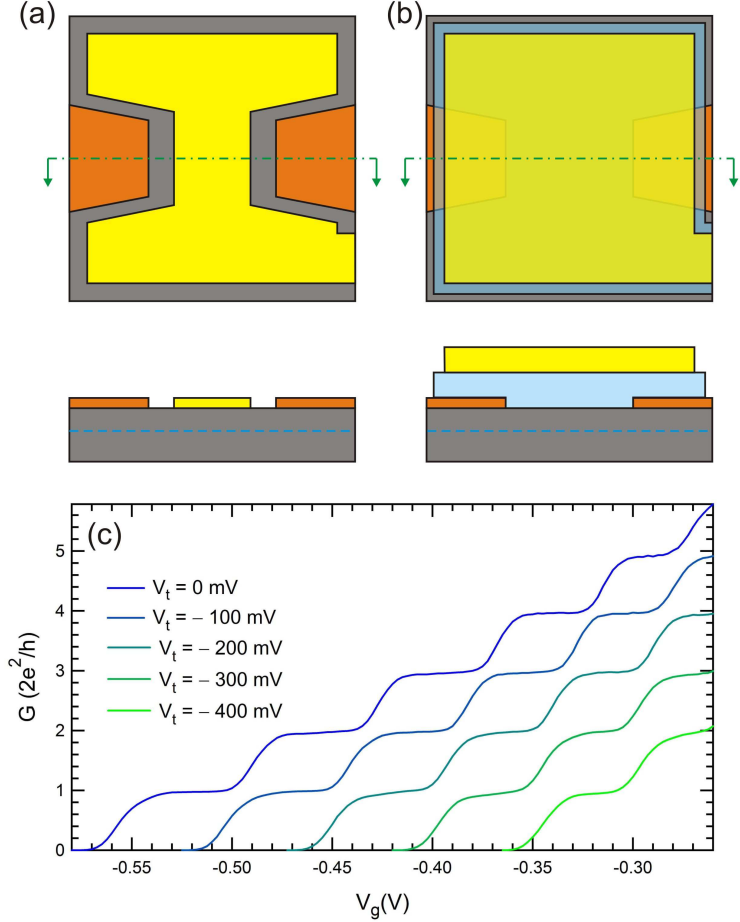}
\caption{Top- and side-view schematics of (a) the bow-tie (BT) and
(b) the polyimide (PI) devices. The side-views are sections along
the green dot-dashed line. The 2DEG (blue dashed line) is located
$90$~nm beneath the heterostructure surface (grey). In both
architectures the QPC gates (orange) define a $300$~nm long,
$500$~nm wide constriction. The top-gate (yellow) controls the 2DEG
density $n$, and is insulated by a $140$~nm thick polyimide layer
(light blue) in the PI device. Gate/insulator structures are drawn
to scale. PI was measured on two separate cool-downs with the QPC
gates trained whilst the top-gate was held at $V_{t} = 0$ and $+
375$~mV, referred to as PI-0 and PI-375, respectively, to enable
$g^{*}_{m}$ measurements for different $n$ ranges. (c) ac
conductance $G$ versus QPC gate voltage $V_{g}$ for five different
$V_{t}$ settings from PI-0. For $V_{g} > -0.25$~V, $G$ rises sharply
due to incomplete gate definition, limiting the density range over
which $g^{*}_{m}$ can be obtained for each 1D subband $m$.}
\end{figure}

We used two different device architectures in this experiment (see
Fig.~1a/b), each fabricated on the same heterostructure and
featuring a pair of QPC gates (orange) biased at $V_{g}$ to define a
1D channel, and a top-gate (yellow) biased at $V_{t}$ to
independently vary $n$. The two devices differ in the location of
the top-gate, allowing us to study how the strength of the 1D
confinement influences $g^{*}_{m}$. The bow-tie (BT) device
(Fig.~1a) has a conformal top-gate with a length of $\sim 60~\mu$m
along the transport direction. The polyimide (PI) device (Fig.~1b)
has a $80 \times 80~\mu$m top-gate separated from the QPC gates by a
$140$~nm polyimide layer. The PI device was measured in two separate
cool-downs, each with different top-gate `training' to give a
slightly different $n$ versus $V_{t}$ characteristic (see
Supplementary Fig.~1). Data is presented for training at $V_{t} = 0$
and $+ 375$~mV, referred to as PI-0 and PI-375 hereafter, providing
three separate `samples' from the two device architectures. A plot
of $n$ versus $V_{t}$ for each sample appears in Supplementary
Fig.~2. The heterostructure used for both devices (Cambridge W0191)
features a $90$~nm deep 2DEG, separated from the modulation doping
layer by a $40$~nm undoped AlGaAs spacer. The 2DEG has a mobility of
$2.7 \times 10^{6}$~cm$^{2}$/Vs at an ungated density of $1.8 \times
10^{11}$~cm$^{-2}$ and temperature of $4$~K. Further device details
appear in the Supplementary Information. The devices were measured
in a dilution refrigerator equipped with a $15$~T superconducting
solenoid and a piezoelectric rotator~\cite{YeohRSI10} for rotating
the sample relative to the applied magnetic field $B$ without the
device temperature exceeding $200$~mK. The density $n$ was measured
with $B$ perpendicular to the 2DEG plane using a Fourier analysis of
the Shubnikov-de Haas oscillations. Measurements of $g^{*}_{m}$ were
obtained with $B$ oriented in-plane and along the QPC axis. To
demonstrate device operation, Fig.~1c shows the ac conductance $G$
versus $V_{g}$ for five different $V_{t}$ spanning the density range
$1.07 - 1.71 \times 10^{11}$~cm$^{-2}$ for PI-0. Conductance
quantization is evident at each $V_{t}$, with pinch-off (i.e., $G =
0$) occurring for smaller $V_{g}$ at more negative $V_{t}$, which
reduces the Fermi energy $E_{F} = \pi\hbar^{2}n/m^{*}$ of the 2DEG
reservoirs adjacent to the QPC. In each case, $G$ rises sharply for
$V_{g} > -0.25$~V due to loss of electrostatic depletion under the
QPC gates. This limits the number of quantized conductance plateaus
observable for each $V_{t}$, and the accessible density range over
which $g^{*}_{m}$ can be obtained for a given 1D subband $m$.

\begin{figure}
\includegraphics[width=16cm]{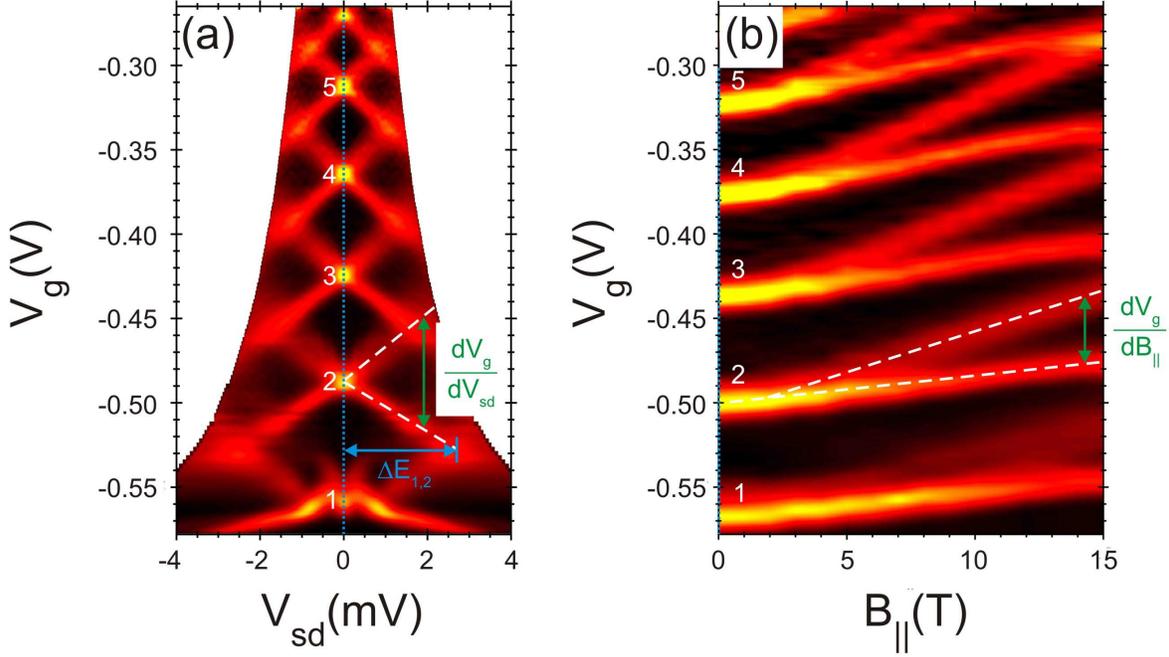}
\caption{ac transconductance $dG/dV_{g}$ vs (a) $V_{g}$ ($y$-axis)
and source-drain bias $V_{sd}$ ($x$-axis) and (b) $V_{g}$ ($y$-axis)
and in-plane magnetic field $B_{\parallel}$ ($x$-axis) for PI-0 with
$V_{t} = 0$~V, corresponding to the left-most (blue) trace in
Fig.~1c. High transconductance (risers in $G$ between plateaus)
appear bright and indicate that a given 1D subband has crossed the
chemical potential $\mu$. The respective 1D subband indices $m$ are
superimposed in both panels. The data in (a) allows us to measure
the lowest 1D subband spacing $\Delta E_{1,2}$ and the
bias-splitting rate $dV_{g}/dV_{sd}$ and (b) the Zeeman splitting
$\Delta E_{z}$ in units of $V_{g}$. We combine the latter two
measurements to obtain the $g^{*}_{m}$ values in Fig.~3, with
$\Delta E_{1,2}$ used to characterize the confining potential in
Fig.~4. The data in (a) has been symmetrized about $V_{sd} = 0$ to
remove an asymmetric background artifact arising from instrumental
issues in the measurement.~\cite{KristensenPRB00}}
\end{figure}

We extract $g^{*}_{m}$ using the method developed by Patel {\it et
al}~\cite{PatelPRB91} to enable direct comparison with the
literature.~\cite{PatelPRB91, ThomasPRL96, MartinAPL08, MartinPRB10}
Two measurements are required to extract $g^{*}_{m}$ at each $n$:
The first is source-drain bias spectroscopy; Fig.~2a shows a
color-map of the ac transconductance $dG/dV_{g}$ against
source-drain bias $V_{sd}$ ($x$-axis) and $V_{g}$ ($y$-axis). The
blue dotted vertical line in Fig.~2a corresponds to the left-most
(blue) trace in Fig.~1c. The bright regions indicate high
transconductance and correspond to the risers between plateaus,
which occur when a 1D subband crosses the source/drain chemical
potential $\mu = E_{F}$. With increasing $V_{sd}$ (i.e., moving
right in Fig.~2a) the source and drain chemical potentials separate
in energy $\mu_{s} - \mu_{d} = eV_{sd}$ producing a bifurcation of
the $V_{sd} = 0$ transconductance maxima (white dashed lines). The
rising/falling bright line corresponds to a given 1D subband
coinciding with $\mu_{d}$ and $\mu_{s}$, respectively. The 1D
subband spacing $\Delta E_{1,2}$ is obtained as $eV_{sd}$ at the
crossing point between the lowest rising line and the second lowest
falling line (blue arrow in Fig.~2a). This provides an important
measure of the `strength' of the transverse confinement at the
center of the QPC; however, as we discuss later, it only provides
partial information about the overall confinement potential
landscape of the QPC. The second measurement is the Zeeman splitting
of the 1D subbands; Fig.~2b shows a color-map of $dG/dV_{g}$ against
in-plane magnetic field $B_{\parallel}$ ($x$-axis) and $V_{g}$
($y$-axis). The blue dotted line corresponds to the left-most (blue)
trace in Fig.~1c. Each 1D subband splits with increasing
$B_{\parallel}$ (white dashed lines); however, this does not
directly yield the Zeeman splitting $\Delta E_{z}$ because the
$y$-axis has units of voltage not energy. To extract the Zeeman
splitting, the splitting rates in Figs.~2a and b are combined, viz:

\begin{equation}
\Delta E_{z} = e[\frac{dV_{g}}{dV_{sd}}]^{-1}\times
\frac{dV_{g}}{dB_{\parallel}} = e\frac{dV_{sd}}{dB_{\parallel}}
\end{equation}

\noindent giving the $g$-factor as $g^{*} = \Delta
E_{z}/\mu_{B}B_{\parallel}$, where $\mu_{B}$ is the Bohr magneton.
The two terms $\frac{dV_{g}}{dV_{sd}}$ and
$\frac{dV_{g}}{dB_{\parallel}}$ are obtained at the same $V_{g}$,
making the confinement potential the same for both contributions to
$E_{z}$, and hence $g^{*}_{m}$. Further details of the analysis
appear in the Supplementary Information.

\begin{figure}
\includegraphics[width=16cm]{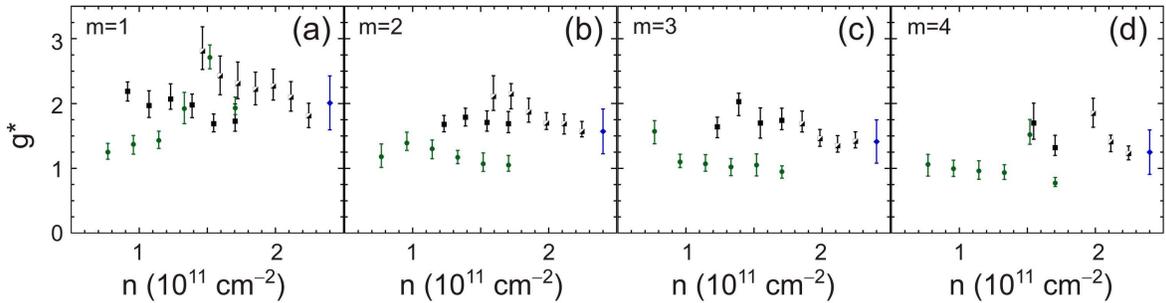}
\caption{Plots of (a) $g^{*}_{1}$, (b) $g^{*}_{2}$, (c) $g^{*}_{3}$
and (d) $g^{*}_{4}$ versus 2DEG density $n$. The black filled
squares, black half-filled squares and green circles correspond to
data from PI-0, PI-375 and BT, respectively. The blue circles/error
bars show the mean and standard deviation for the set of data points
in each of the four panels.}
\end{figure}

We now examine how $g^{*}_{m}$ evolves with density $n$ for the
lowest four 1D subbands, with Fig.~3a-d presenting $g^{*}_{1}$,
$g^{*}_{2}$, $g^{*}_{3}$ and $g^{*}_{4}$ versus $n$ for PI-0, PI-375
and BT. At each density $g^{*}_{m}$ is obtained from an individually
measured pair of source-drain bias and field plots similar to those
in Fig.~2. The blue circle and error bar in each panel of Fig.~3
represents the mean and standard deviation for the full set of data
presented in that panel; comparing these for panels a-d, $g^{*}_{m}$
clearly increases with decreasing $m$ on average (see also
Supplementary Fig.~3), consistent with previous
studies~\cite{PatelPRB91, ThomasPRL96, MartinAPL08, MartinPRB10}.
The density-dependence of $g^{*}_{m}$ is complex and evolves with
$m$. We start first at $m \geq 2$. Considering each individual
device on its own for a moment, in each panel (b-d) we see that
$g^{*}_{m}$ mostly increases with decreasing $n$, as one would
expect for exchange interactions.~\cite{JanakPR69} However,
considering the full three device data-set in each panel (b-d) there
is no clear trend with density. At $m = 1$ a distinct difference in
the density dependencies for the PI and BT devices emerges
(Fig.~3a): as $n$ is reduced we observe increasing $g^{*}_{1}$ for
both PI samples but decreasing $g^{*}_{1}$ for BT. Note also the
lack of overlap in the individual $g^{*}_{1}$ versus $n$ behavior
for PI-0 and PI-375 in the common density range $1.5 - 1.7 \times
10^{11}$~cm$^{-2}$. The findings above clearly point to a
relationship between $g^{*}$ and $n$ that depends also on other
parameters. It is worth noting that there is evidence for disorder
effects in BT, as discussed further below, which are more prevalent
than for PI-0/PI-375. One possibility that we cannot rule out at
this stage is that the inherent disorder potential may also have an
effect in $g^{*}(n)$. That said, given the identical heterostructure
and QPC gate pattern for PI-0, PI-375 and BT, a natural expectation
is that the difference is due to the top-gate; hence the logical
next step is to consider the influence of the QPC confinement
potential.

\begin{figure}
\includegraphics[width=16cm]{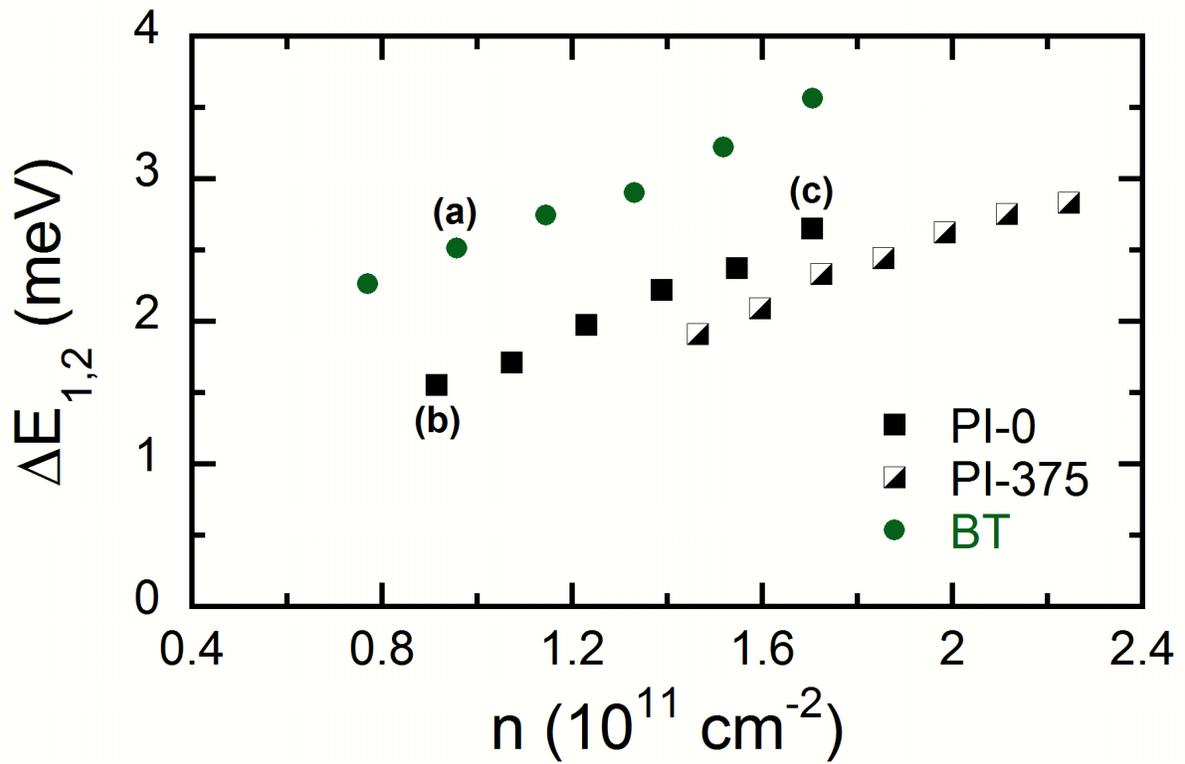}
\caption{Plot of the lowest 1D subband spacing $\Delta E_{1,2}$ vs
density $n$ for PI-0, PI-375 and BT. The letters (a-c) indicate the
points corresponding to the data presented in Fig.~7.}
\end{figure}

Precise knowledge of QPC confinement potential is difficult, it not
only depends on gate bias, but also on the heterostructure doping
profile and the self-consistent redistribution of charge in the
2DEG.~\cite{LauxSurfSci88, HirosePRL03, RejecNat06} The QPC is most
commonly treated as a saddle-point potential,~\cite{ButtikerPRB90}
and while more sophisticated self-consistent models yield a flat
bottomed parabola for the transverse potential,~\cite{LauxSurfSci88}
a simple parabola is usually sufficient, particularly in the small
$m$ limit.~\cite{WeiszPRB89, KardynalPRB97, LeeJAP06}. In a 1D
parabolic well, the energy level spacing is directly tied to the
curvature; hence the 1D subband spacing $\Delta E_{m,m+1}$ is
commonly used as a metric for the 1D confinement strength in
QPCs.~\cite{ThomasAPL95, KristensenJAP98, LiangPRB99, LeeJAP06,
KoopJSNM07, HewPRL08} Figure~4 shows the measured $\Delta E_{1,2}$
versus $n$ for all three samples; we focus solely on $\Delta
E_{1,2}$ due to our interest in $g^{*}_{1}$ regarding both
spintronic applications and the $0.7$ anomaly. In each case $\Delta
E_{1,2}$ decreases as $n$ is reduced, indicating a softening 1D
confinement potential, consistent with earlier studies using similar
device architectures.~\cite{LeeJAP06, HewPRL08}

\begin{figure}
\includegraphics[width=\linewidth]{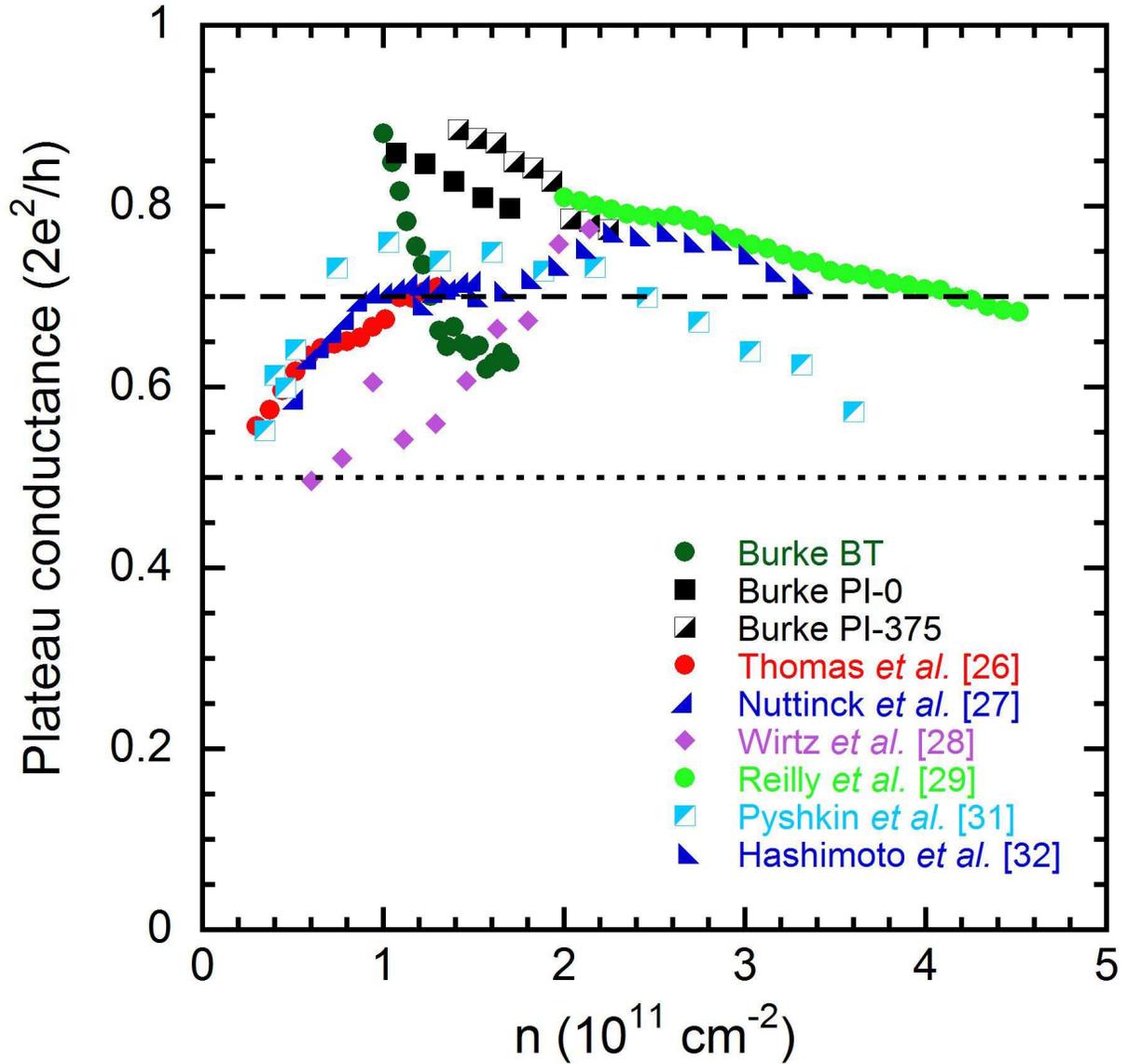}
\caption{The precise conductance $G$ of the $0.7$ anomaly vs density
$n$ for Refs.~\cite{ThomasPRB00, NuttinckJJAP00, WirtzPRB02,
ReillyPRB01, PyshkinPRB00, HashimotoJJAP01} and our data from PI-0,
PI-375 and BT in Fig.~6. The devices in Refs.~\cite{ThomasPRB00,
ReillyPRB01} have a midline gate similar to BT (solid circles) and
in Ref.~\cite{PyshkinPRB00} has a polyimide-insulated top-gate
similar to PI (half filled squares). No simple link between the
location/evolution of the $0.7$ plateau and $n$ is evident
suggesting the $0.7$ anomaly is heavily dependent on how the QPC is
defined.}
\end{figure}

Comparing Figs.~3a and 4, an interesting but complex connection
between $g^{*}_{1}$, $n$ and $\Delta E_{1,2}$ emerges. Considering
PI-0 and PI-375 alone first; their $\Delta E_{1,2}$ versus $n$
trends almost overlap and differ by at most $16 \%$ in the common
density range $n = 1.4 - 1.8 \times 10^{11}~$cm$^{-2}$, indicating a
similar 1D confinement strength. Despite this, $g^{*}_{1}$ differs
markedly ($\sim 1.8$ for PI-0 and $\sim 2.5$ for PI-375). Turning to
PI-0 and BT over the wider density range $n = 0.8 - 1.8 \times
10^{11}~$cm$^{-2}$, the different top-gate implementation results in
a $\Delta E_{1,2}$ that is consistently $\sim 160\%$ larger for BT
than PI-0, indicating a harder, more square-well like confinement
potential for BT. Yet, as Fig.~3a shows, not only can $g^{*}_{1}$
differ markedly between PI-0 and BT at a given density, but in the
former we observe increasing $g^{*}_{1}$ and the latter decreasing
$g^{*}_{1}$ with decreasing $n$. Finally, at $n = 1.7 \times
10^{11}~$cm$^{-2}$ where data exists for all three samples, we find
$g^{*}_{1}(\textrm{PI-375}) > g^{*}_{1}(\textrm{BT}) >
g^{*}_{1}(\textrm{PI-0})$ whereas $\Delta E_{1,2}(\textrm{BT}) >>
\Delta E_{1,2}(\textrm{PI-0}) \sim \Delta E_{1,2}(\textrm{PI-375})$.
Hence while Koop {\it et al}~\cite{KoopJSNM07} suggest a clear and
direct correlation between $g^{*}$ and $\Delta E_{1,2}$, our data
suggests the connection is much more subtle. The subband spacing
$\Delta E_{1,2}$ is only sensitive to the transverse quasi-parabolic
1D potential at the center of the QPC. As such, $\Delta E_{1,2}$ is
a limited metric of the overall shape of the QPC potential
landscape, which depends on length, width,
density~\cite{LauxSurfSci88,IqbalArxiv12} and realistically, the
inherent disorder potential~\cite{NixonPRB91}. Our data suggests
that $g^{*}_{1}$ is heavily dependent on the overall QPC potential,
presumably via its influence on exchange interactions within the
QPC. Spin density functional theory (SDFT) calculations also point
to exchange effects being very sensitive to the precise geometry of
the QPC as the device approaches pinch-off.~\cite{RejecNat06,
AkisJPCM08, BerggrenJPCM08}. The reduced variability in $g^{*}$ with
increasing $m \geq 2$ may be due to improved screening arising from
the higher electron density within the QPC (even with fixed $n$).

\begin{figure}
\includegraphics[width=10cm]{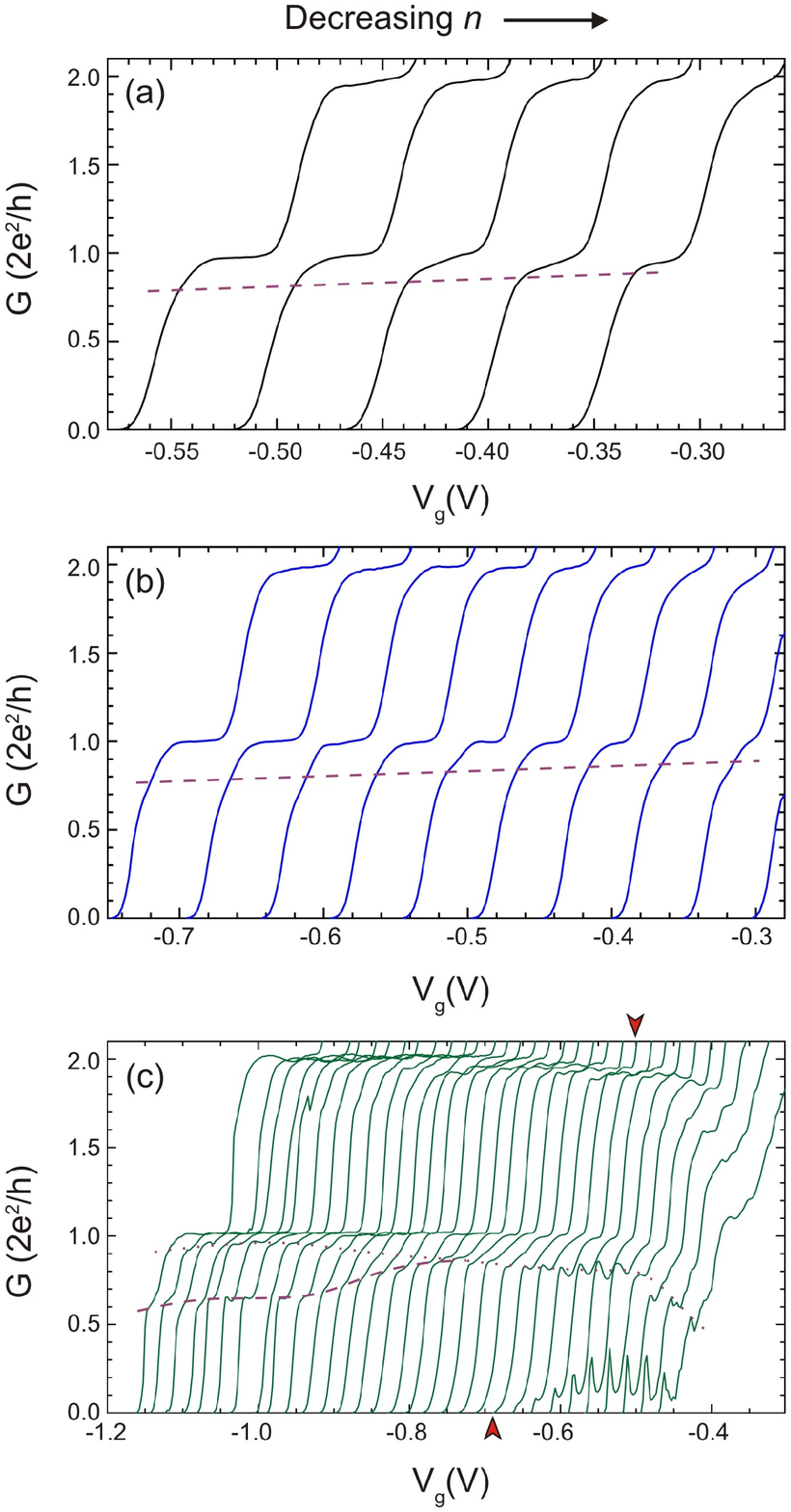}
\caption{Conductance $G$ versus QPC gate voltage $V_{g}$ as top-gate
voltage $V_{t}$ is changed for (a) PI-0, (b) PI-375 and (c) BT. The
$V_{t}$ ranges and increments are (a) $0$ (left) to $-400$~mV
(right) in steps of $100$~mV, (b) $+375$ (left) to $-375$~mV (right)
in steps of $75$~mV, and (c) $0$ (left) to $-210$~mV (right) in
steps of $7$~mV. The purple dashed lines are guides to the eye
highlighting the evolution of anomalous plateau-like structures at
$G < G_{0}$ with density $n$. The red arrows at top/bottom of (c)
indicate the right-most trace from which $g^{*}_{m}$ data is
extracted.}
\end{figure}

Considering the applied implications first, when using a QPC as a
spin injector/detector, one commonly applies a magnetic field to
break the 1D subband degeneracy, and operates the QPC at $G <
0.5G_{0}$ so that transmission is dominated by the $1\downarrow$
subband.~\cite{PotokPRL02, FolkSci03} This makes maximizing
$g^{*}_{1}$ of prime importance in reducing the magnetic field
required for operation. As Fig.~3a shows, we can achieve $g^{*}_{1}$
as large as $2.8$ approaching the high values obtained in InGaAs
QPCs,~\cite{MartinAPL08} and exceeding the $0.75 - 1.5$ typically
reported for GaAs QPCs with the field applied in the plane of the
2DEG.~\cite{PatelPRB91, ThomasPRL96, KoopJSNM07, FrolovPRL09} We
note that $g^{*} \sim 4$ was recently reported for a GaAs QPC with
the field oriented perpendicular to the 2DEG by R\"{o}ssler {\it et
al}.~\cite{RosslerNJP11} Although substantially higher $g^{*}$
values can be obtained in QPCs with the field perpendicular to the
2DEG,~\cite{MartinPRB10} due to the strong confinement in the
heterostructure growth direction~\cite{KowalskiPRB94} and exchange,
this comes with associated problems of cyclotron curvature, and at
higher fields $B > 1-2$~T, Landau quantization. These can be
problematic for spintronic applications, such that in-plane fields
are more commonly used; here the tight confinement of the 2DEG
allows fields exceeding $10$~T to be applied before cyclotron issues
arise, which more than compensates for the lower $g^{*}$ obtained
for in-plane fields. An additional benefit of using an in-plane
field to break the spin-degeneracy is that an independently variable
and smaller perpendicular field component can be used to `steer' a
ballistic electron beam into a spin-polarized collector
QPC.~\cite{FolkSci03, RokhinsonPRL04} The trends in Fig.~3a are also
important, because in addition to $g^{*}_{1}$ values as high as
$2.8$ we see values as low as $1.25$. Hence the key to achieving and
maintaining high $g^{*}_{1}$ in QPCs is very careful management of
the QPC's confinement potential and local electrostatic environment
(e.g., 2DEG density).

The evolution of $g^{*}_{1}$ and $\Delta E_{1,2}$ with $n$ in
Figs.~3a and 4 also provides an opportunity to address the conflict
in the literature regarding the density dependence of the $0.7$
anomaly. Briefly reviewing the various observations: the $0.7$
plateau was reported to gradually fall towards $0.5 G_{0}$ with {\it
decreasing} $n$ in three devices: a modulation-doped midline-gated
QPC~\cite{ThomasPRB00} (similar to BT), a modulation-doped
back-gated QPC~\cite{NuttinckJJAP00} and a modulation-doped QPC
where $n$ was changed using illumination/hydrostatic
pressure~\cite{WirtzPRB02}. In contrast, the {\it opposite}
dependence (i.e., $0.7$ falls to $0.5 G_{0}$ with increasing $n$) is
observed in two other devices: an undoped QPC with a positively
biased bow-tie top-gate~\cite{ReillyPRB01} and a modulation-doped
QPC with a mid-line gate~\cite{LeeJAP06}. Finally, a $0.7$ plateau
that fell towards $0.5 G_{0}$ with {\it both} increasing and
decreasing $n$ was reported for an undoped QPC with a
polyimide-insulated top-gate,~\cite{PyshkinPRB00} and a $0.7$
plateau that is strong at low and high $n$ and which weakens whilst
rising to $0.8 G_{0}$ at intermediate $n$ was reported for a
modulation-doped back-gated QPC~\cite{HashimotoJJAP01}. A summary of
these results is presented in Fig.~5,~\footnote{Results by Lee {\it
et al}~\cite{LeeJAP06} are omitted from Fig.~5 as we are unable to
precisely determine $n$ values for this data.} where we plot the $G$
at which the $0.7$ plateau appears against density $n$ for the data
in Refs.~\cite{ThomasPRB00, NuttinckJJAP00, WirtzPRB02, ReillyPRB01,
PyshkinPRB00, HashimotoJJAP01}, along with our data from Fig.~6.
There is no straightforward link between the location/evolution of
the $0.7$ anomaly and density in Fig.~5, instead the behavior
appears highly device-dependent. The QPC confinement potential is
also the crux of this problem, as we now show.

\begin{figure}
\includegraphics[width=16cm]{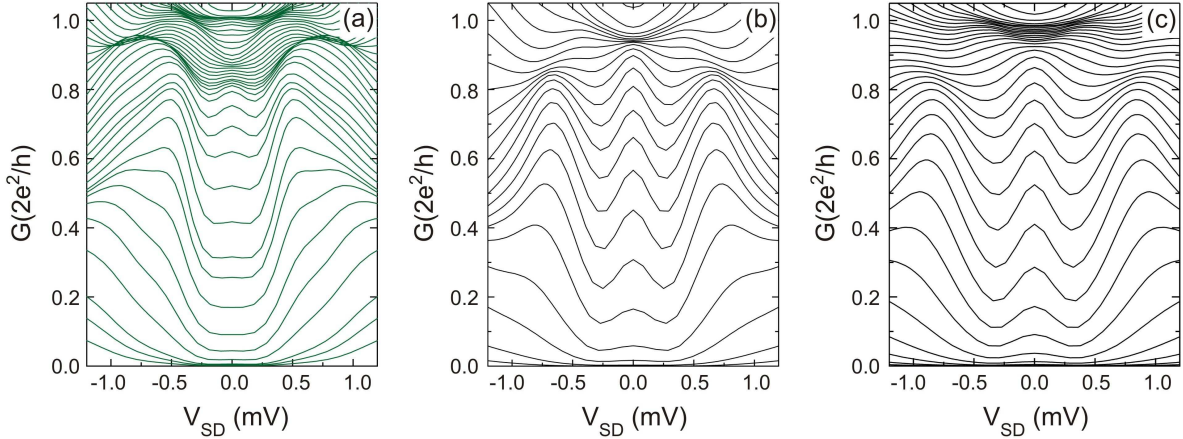}
\caption{Differential conductance $G^{\prime}(V_{sd})$ vs
source-drain bias $V_{sd}$ for a range of $V_{g}$ for (a) BT at
$V_{t} = -120$~mV, (b) PI-0 at $V_{t} = -500$~mV and (c) $0$~mV. In
each case $T < 100$~mK and $B = 0$, with (a) $n = 0.95 \times
10^{11}$~cm$^{-2}$ and $\Delta E_{1,2} = 2.52$~meV (b) $n = 0.92
\times 10^{11}$~cm$^{-2}$ and $\Delta E_{1,2} = 1.56$~meV, and (c)
$n = 1.70 \times 10^{11}$~cm$^{-2}$ and $\Delta E_{1,2} = 2.66$~meV.
The width and amplitude of the zero-bias peak (ZBP) is relatively
constant with $n$ for PI-0 and PI-375. ZBP suppression is a common
feature for all $n$ in BT. The data in each panel has been
symmetrized about $V_{sd} = 0$ to remove an asymmetric background
artifact arising from instrumental issues in the
measurement.~\cite{KristensenPRB00}}
\end{figure}

Figure~6a-c shows $g$ versus $V_{g}$ as a function of $V_{t}$ for
PI-0, PI-375 and BT, respectively. The dashed purple lines highlight
the evolution of the $0.7$ anomaly with $n$. Before considering the
experimental data itself, we briefly digress to consider some
predictions based on the density-dependent spin-gap model introduced
by Reilly {\it et al},~\cite{ReillyPRL02, ReillyPRB05} since these
will be vital to the discussion. The one free parameter in this
model, the opening rate $\gamma = d\Delta
E_{\uparrow\downarrow}/dV_{g}$ of the spin-gap $\Delta
E_{\uparrow\downarrow}$ with QPC gate bias $V_{g}$, is suggested to
be linked to the potential mismatch between the 1D channel and 2D
reservoirs.~\cite{ReillyPRL02, ReillyPRB05} This mismatch is
essentially a mode-matching effect,~\cite{GlazmanJETP88,
SzaferPRL89} and should be dependent on the 1D confinement strength,
i.e., $\Delta E_{1,2}$.~\cite{HirosePRL03, LindelofJPCM08, WuPRB12}
Considering Fig.~3(b) of Ref.~\cite{ReillyPRB05}, the prediction is
that for increasing $\gamma$ the $0.7$ anomaly will move lower in
$G$ and become more pronounced. If we take the simplest assumption
$\gamma \propto \Delta E_{1,2}$,~\cite{ReillyPRB05} two behaviors
are expected given the $\Delta E_{1,2}$ data in Fig.~4: a) the $0.7$
anomaly should be weak and at higher $G$ for PI-0 and PI-375 and be
more pronounced and at lower $G$ in BT, and b) in each case the
$0.7$ anomaly should weaken and tend to rise in $G$ with decreasing
$n$.

Considering the experimental data now; for both PI-0 and PI-375 we
observe a weak inflection at relatively high $G < G_{0}$
(Fig.~6(a/b)). This inflection moves to higher $G$ with decreasing
$n$ in both cases, consistent with the corresponding reduction in
$\Delta E_{1,2}$ in Fig.~4. Because the plateau appears as a weak
inflection, its weakening with decreasing $n$ is unfortunately
difficult to distinguish. However, we note that a similar rise in
conductance of the $0.7$ anomaly is observed with reduced $\Delta
E_{1,2}$ by Liang {\it et al} (see Fig.~2 of
Ref.~\cite{LiangPRB99}), and here the associated weakening of the
$0.7$ plateau is more visible. It is also unclear whether the rise
in the $0.7$ feature with decreasing $n$ is linear; a careful look
at Fig.~6(b) suggests it may not be (see also Fig.~6(c)). This would
mean that the assumption $\gamma \propto \Delta E_{1,2}$ may only be
approximate to the true functional relationship between $\gamma$ and
$\Delta E_{1,2}$. For the BT device, the $0.7$ anomaly starts as a
clear plateau at $G \sim 0.6 G_{0}$ in the high $n$ limit. The
plateau rises in conductance and weakens with decreasing $n$,
ultimately merging in the low $n$ limit with an additional anomalous
feature dropping down from higher $G$ with decreasing $n$, as
highlighted by the dotted purple line. The observation of an
additional plateau above the $0.7$ anomaly is relatively
common~\cite{CrookSci06, ThomasPRB00, ThomasPM98, LeeJAP06} as are
small `bumps' on the $0.7$ anomaly at higher
density.~\cite{ReillyPRL02, LeeJAP06}. The appearance of this
structure in BT rather than PI is not unexpected -- higher $\Delta
E_{1,2}$ is associated with a more sharply defined 1D channel, and
this should enhance resonant features in the conductance, as
discussed by Kirczenow.~\cite{KirczenowPRB89} While the BT data
should be considered with care as disorder effects can modify the
behaviour of the $0.7$ anomaly, the $0.7$ feature behavior we
observe for BT is entirely consistent with expectations based on the
density-dependent spin-gap model~\cite{ReillyPRB05} and the behavior
observed for PI-0/PI-375: The $0.7$ plateau in BT is stronger and
appears at a lower conductance than for the PI samples, consistent
with the much higher $\Delta E_{1,2}$ in Fig.~4. The $0.7$ plateau
rises and weakens with decreasing $n$ in each case also, following
expectations based on Fig.~4 and the density-dependent spin-gap
model.

One possible explanation for our observations, in particular the
link between the $0.7$ anomaly behavior and $\Delta E_{1,2}$, but
also the sensitivity of $g^{*}_{1}$, is the formation of a
quantum-dot-like localized charge state within the QPC due to the
1D-2D mismatch at the QPC openings. There is strong experimental
evidence that this can occur in QPCs including observations of
Kondo-like behavior~\cite{CronenwettPRL02, SfigakisPRL08,
SarkozyPRB09, ChenPRB09, KomijaniEPL10, KlochanPRL11, WuPRB12}, Fano
resonances~\cite{YoonPRL07, YoonPRB09, YoonPRX12} and
Fabry-P\'{e}rot oscillations~\cite{LindelofJPCM08} on the integer
conductance plateaus by other authors.\footnote{We also observe weak
Fabry-P\'{e}rot-like structure along $V_{sd} = 0$ in some of our
source-drain bias plots (see Supplementary Fig.~4); it is much
stronger in Ref.~\cite{LindelofJPCM08}, presumably due to the
stronger 1D confinement ($\Delta E_{1,2} \sim 5$~meV).} Strong
exchange effects that are very sensitive to the size, shape and
symmetry are a well-known feature of ultra-small quantum
dots.~\cite{KouwenhovenRPP01, HansonRMP07} We propose that
variations in the size, shape and symmetry of a quantum-dot-like
localized charge state within a QPC might similarly affect the
exchange interaction in QPCs, and thereby affect both the $g$-factor
and $0.7$ plateau behavior. Under this explanation, the $g$-factor
and its sensitivity to confinement should decrease with increasing
subband index. Accordingly, the magnitude of $g^{*}$ decreases with
increasing $m$ in Fig.~4; however, a reduced sensitivity to
confinement is not as apparent.

The mention of Kondo-like behavior above naturally leads to the
question of the zero-bias peak in the differential conductance
$G^{\prime}(V_{sd})$, a feature first discussed in detail by
Cronenwett {\it et al},~\cite{CronenwettPRL02} and commonly observed
in QPCs.~\cite{MicolichJPCM11} The role played by Kondo physics in
QPCs is heavily debated;~\cite{MicolichJPCM11} some suggest that
Kondo physics drives the weakening of the $0.7$ plateau in the low
$T$ limit,~\cite{CronenwettPRL02, MeirPRL02} others argue for the
$0.7$ plateau and Kondo-like physics in QPCs being separate and
distinct effects.~\cite{SfigakisPRL08} In Fig.~7 we plot
$G^{\prime}(V_{sd})$ versus $V_{sd}$ for BT and PI-0, with $V_{t}$
settings chosen to best isolate the effect of differences in $\Delta
E_{1,2}$ and $n$ (the corresponding data points are indicated (a-c)
in Fig.~4). PI-0 shows a clear zero-bias peak (ZBP) over the entire
range $0 < G < G_{0}$, this ZBP behavior is consistent with most
previous reports.~\cite{CronenwettPRL02, SfigakisPRL08,
KomijaniEPL10, KlochanPRL11, MicolichJPCM11} The ZBP amplitude and
width for a given $G$ are relatively independent of density
(Fig.~7b/c); similar behavior is found for PI-375. In comparison,
the ZBP for BT is heavily suppressed (Fig.~7a); whilst evident as a
smaller amplitude peak for $0.4 G_{0} < G < 0.8 G_{0}$ it vanishes
in the limits $G \rightarrow 0$ and $G \rightarrow G_{0}$. This
behavior also holds qualitatively as density is varied, but the
amplitude, width and suppression vary slightly; an in-depth study
will be presented elsewhere. Thermally-induced suppression of the
Kondo process, and hence the ZBP, occurs as the temperature $T$
increases relative to the Kondo temperature
$T_{K}$.~\cite{GoldhaberNat98} In QPCs, this typically occurs for $T
> 0.5$~K.~\cite{CronenwettPRL02, SfigakisPRL08, KomijaniEPL10}
However, in our experiment $T$ is fixed at $< 100$~mK; hence it must
be $T_{K}$ that differs between PI-0 and BT, with $T_{K}^{BT} <
T_{K}^{PI}$. This difference in $T_{K}$ is not unexpected. In
quantum dots, $T_{K}$ depends sensitively on the charging energy
$U$, the bound-state energy $\epsilon_{0}$ and the coupling $\Gamma$
to the reservoirs.~\cite{GoldhaberNat98} These can be independently
tuned in dots,~\cite{GoldhaberNat98} but not for a localized charge
state within a QPC. Simulations predict $\Gamma$ to be particularly
sensitive to QPC potential,~\cite{HirosePRL03} but the influence of
other factors such as QPC length, width, and most notably 1D-2D
mismatch, i.e., 1D confinement strength $\Delta E_{1,2}$, are
unknown. Note however that the data in Fig.~6 is consistent with a
Kondo-like scenario:~\cite{CronenwettPRL02, MeirPRL02, HirosePRL03}
for BT where the ZBP is suppressed, the $0.7$ plateau is more
evident and at a lower conductance than for PI, where the ZBP is
stronger. The behavior in Fig.~7 cannot be tied directly to $\Delta
E_{1,2}$ or $n$ alone; we suggest that it instead depends on the
precise nature of the QPC confinement potential. The change in the
ZBP we observe may indicate a change in the coupling of a single
localized state to the reservoirs, or potentially to the emergence,
loss, or interaction of multiple localized states within the QPC as
$G$ is driven from $G_{0}$ to $0$.~\cite{RejecNat06, MeirPC}

In conclusion, we have studied the dependence of the 1D Land\'{e}
$g$-factor $g^{*}$ on density in QPCs with two different top-gate
architectures. We obtain $g^{*}$ values for the lowest 1D subband of
up to $2.8$, approaching the high values obtained in InGaAs/InP
QPCs~\cite{MartinAPL08}, and significantly exceeding previously
reported values for the in-plane $g$-factor of AlGaAs/GaAs
QPCs.~\cite{PatelPRB91, ThomasPRL96, CronenwettPRL02, KoopJSNM07,
FrolovPRL09} Careful management of the QPC's confinement potential
appears key to obtaining high $g^{*}_{1}$. This has important
implications for using QPCs in spintronic applications. The
appearance of the $0.7$ plateau is strongly linked to 1D confinement
potential, explaining the conflicting density dependencies reported
in the literature.~\cite{ThomasPRB98, ThomasPRB00, NuttinckJJAP00,
PyshkinPRB00, HashimotoJJAP01, ReillyPRB01, WirtzPRB02, ReillyPRL02,
ReillyPRB05, LeeJAP06} In particular, the $0.7$ anomaly behavior in
our devices is consistent with predictions made using the
density-dependent spin-gap model~\cite{ReillyPRL02, ReillyPRB05}
with the one free parameter taken as directly proportional to the
lowest 1D subband spacing $\Delta E_{1,2}$.

{\bf Supporting Information.} Extended details of methods used,
device characterization data and additional supporting data. This
material is available free of charge via the Internet at
http://pubs.acs.org.

{\bf Corresponding author.} *E-mail:
adam.micolich@nanoelectronics.physics.unsw.edu.au

\acknowledgement

This work was funded by the Australian Research Council (ARC)
through the Discovery Projects Scheme. APM acknowledges an ARC
Future Fellowship (FT0990285). ARH acknowledges an ARC Professorial
Fellowship. IF and DAR acknowledge financial support from the EPSRC.
We thank T.P. Martin, U. Z\"{u}licke, D.J. Reilly and Y. Meir for
helpful discussions.


\begin{thebibliography}:

\bibitem{AwschalomNP07} Awschalom, D.D.; Flatt\'{e}, M.E. {\it Nature Physics} {\bf 2007}, {\it 3}, 153-159.

\bibitem{AwschalomPhys09} Awschalom, D.D.; Samarth, N. {\it Physics} {\bf 2009}, {\it 2}, 50-54.

\bibitem{PotokPRL02} Potok, R.M.; Folk, J.A.; Marcus, C.M.; Umansky, V. {\it Phys. Rev. Lett.} {\bf 2002}, {\it 89}, 266602.

\bibitem{FolkSci03} Folk, J.A.; Potok, R.M.; Marcus, C.M.; Umansky, V. {\it Science} {\bf 2003}, {\it 299}, 679-682.

\bibitem{DebrayNN09} Debray, P.; Rahman, S.M.S.; Wan, J.; Newrock, R.S.; Cahay, M.; Ngo, A.T.; Ulloa, S.E.; Herbert, S.T.; Muhammad, M.; Johnson, M. {\it Nature Nanotech.} {\bf 2009}, {\it 4}, 759-764.

\bibitem{FrolovPRL09} Frolov, S.M.; Venkatesan, A.; Yu, W.; Folk, J.A.; Wegscheider, W. {\it Phys. Rev. Lett.} {\bf 2009}, {\it 102}, 116802.

\bibitem{FrolovNat09} Frolov, S.M.; L\"{u}scher, S.; Yu, W.; Ren, Y.; Folk, J.A.; Wegscheider, W. {\it Nature} {\bf 2009}, {\it 458}, 868-871.

\bibitem{vanWeesPRL88} van Wees, B.J.; van Houten, H.; Beenakker, C.W.J.; Williamson, J.G.; Kouwenhoven, L.P.; van der Marel, D.; Foxon, C.T. {\it Phys. Rev. Lett.} {\bf 1988}, {\it 60}, 848-850.

\bibitem{WharamJPC88} Wharam, D.A.; Thornton, T.J.; Newbury, R.; Pepper, M.; Ahmed, H.; Frost, J.E.F.; Hasko, D.G.; Peacock, D.C.; Ritchie, D.A.; Jones, G.A.C. {\it J. Phys. C} {\bf 1988}, {\it 21}, L209-L214.

\bibitem{ThomasPRL96} Thomas, K.J.; Nicholls, J.T.; Simmons, M.Y.; Pepper, M.; Mace, D.R.; Ritchie, D.A. {\it Phys. Rev. Lett.} {\bf 1996}, {\it 77}, 135-138.

\bibitem{MicolichJPCM11} Micolich, A.P. {\it J. Phys.: Condens. Matter} {\bf 2011}, {\it 23}, 443201.

\bibitem{AuslaenderSci05} Auslaender, O.M.; Steinberg, H.; Yacoby, A.; Tserkovnyak, Y.; Halperin, B.I.; Baldwin, K.W.; Pfeiffer, L.N.; West, K.W. {\it Science} {\bf 2005}, {\it 308}, 88-92.

\bibitem{HewPRL08} Hew, W.K.; Thomas, K.J.; Pepper, M.; Farrer, I.; Anderson, D.; Jones, G.A.C.; Ritchie, D.A. {\it Phys. Rev. Lett.} {\bf 2008}, {\it 101}, 036801; {\it Phys. Rev. Lett.} {\bf 2009}, {\it 102}, 056804.

\bibitem{DeshpandeNat10} Deshpande, V.V.; Bockrath, M.; Glazman, L.I.; Yacoby, A. {\it Nature} {\bf 2010}, {\it 464}, 209-215.

\bibitem{PatelPRB91} Patel, N.K.; Nicholls, J.T.; Mart\'{\i}n-Moreno, L.; Pepper, M.; Frost, J.E.F.; Ritchie, D.A.; Jones, G.A.C. {\it Phys. Rev. B} {\bf 1991}, {\it 44}, 10973-10975.

\bibitem{JanakPR69} Janak, J.F. {\it Phys. Rev.} {\bf 1969}, {\it 178}, 1416-1418.

\bibitem{ThomasPRB98} Thomas, K.J.; Nicholls, J.T.; Appleyard, N.J.; Simmons, M.Y.; Pepper, M.; Mace, D.R.; Tribe, W.R.; Ritchie, D.A. {\it Phys. Rev. B} {\bf 1998}, {\it 58}, 4846-4852.

\bibitem{MartinAPL08} Martin, T.P.; Szorkovszky, A.; Micolich, A.P.; Hamilton, A.R.; Marlow, C.A.; Linke, H.; Taylor, R.P.; Samuelson, L. {\it Appl. Phys. Lett.} {\bf 2008}, {\it 93}, 012105.

\bibitem{MartinPRB10} Martin, T.P.; Szorkovszky, A.; Micolich, A.P.; Hamilton, A.R.; Marlow, C.A.; Taylor, R.P.; Linke, H.; Xu, H.Q. {\it Phys. Rev. B} {\bf 2010}, {\it 81}, 041303.

\bibitem{CronenwettPRL02} Cronenwett, S.M.; Lynch, H.J.; Goldhaber-Gordon, D.; Kouwenhoven, L.P.; Marcus, C.M.; Hirose, K.; Wingreen, N.S.; Umansky, V. {\it Phys. Rev. Lett.} {\bf 2002} {\it 88}, 226805.

\bibitem{KoopJSNM07} Koop, E.J.; Lerescu, A.I.; Liu, J.; van Wees, B.J.; Reuter, D.; Wieck, A.D.; van der Wal, C.H. {\it J. Supercond. Nov. Magn.} {\bf 2007}, {\it 20}, 433-441.

\bibitem{RosslerNJP11} R\"{o}ssler, C.; Baer, S.; de Wiljes, E.; Ardelt, P.-L.; Ihn, T.; Ensslin, K.; Reichl, C.; Wegscheider, W. {\it New J. Phys.} {\bf 2011}, {\it 13}, 113006.

\bibitem{SchapersAPL07} Sch\"{a}pers, T.; Guzenko, V.A.; Hardtdegen, H. {\it Appl. Phys. Lett.} {\bf 2007}, {\it 90}, 122107.

\bibitem{ReillyPRL02} Reilly, D.J.; Buehler, T.M.; O'Brien, J.L.; Hamilton, A.R.; Dzurak, A.S.; Clark, R.G.; Kane, B.E.; Pfeiffer, L.N.; West, K.W. {\it Phys. Rev. Lett.} {\bf 2002}, {\it 89}, 246801.

\bibitem{ReillyPRB05} Reilly, D.J. {\it Phys. Rev. B} {\bf 2005}, {\it 72}, 033309.

\bibitem{ThomasPRB00} Thomas, K.J.; Nicholls, J.T.; Pepper, M.; Tribe, W.R.; Simmons, M.Y.; Ritchie, D.A. {\it Phys. Rev. B} {\bf 2000}, {\it 61}, 13365-13368.

\bibitem{NuttinckJJAP00} Nuttinck, S.; Hashimoto, K.; Miyashita, S.; Saku, T.; Yamamoto, Y.; Hirayama, Y. {\it Jpn. J. Appl. Phys.} {\bf 2000}, {\it 39}, L655-L657.

\bibitem{WirtzPRB02} Wirtz, R.; Newbury, R.; Nicholls, J.T.; Tribe, W.R.; Simmons, M.Y.; Pepper, M. {\it Phys. Rev. B} {\bf 2002}, {\it 65}, 233316.

\bibitem{ReillyPRB01} Reilly, D.J.; Facer, G.R.; Dzurak, A.S.; Kane, B.E.; Clark, R.G.; Stiles, P.J.; Hamilton, A.R., O'Brien, J.L.; Lumpkin, N.E.; Pfeiffer, L.N.; West, K.W. {\it Phys. Rev. B} {\bf 2001}, {\it 63}, 121311.

\bibitem{LeeJAP06} Lee, H.-M.; Muraki, K.; Chang, E.Y.; Hirayama, Y. {\it J. Appl. Phys.} {\bf 2006}, {\it 100}, 043701.

\bibitem{PyshkinPRB00} Pyshkin, K.S.; Ford, C.J.B.; Harrell, R.H.; Pepper, M.; Linfield, E.H.; Ritchie, D.A. {\it Phys. Rev. B} {\bf 2000}, {\it 62}, 15842-15850.

\bibitem{HashimotoJJAP01} Hashimoto, K.; Miyashita, S.; Saku, T.; Hirayama, Y. {\it Jpn. J. Appl. Phys.} {\bf 2001}, {\it 40}, 3000-3002.

\bibitem{YeohRSI10} Yeoh, L.A.; Srinivasan, A.; Martin, T.P.; Klochan, O.; Micolich, A.P.; Hamilton, A.R. {\it Rev. Sci. Instrum.} {\bf 2010}, {\it 81}, 113905.

\bibitem{KristensenPRB00} Kristensen, A.; Bruus, H.; Hansen, A.E.; Jensen, J.B.; Lindelof, P.E.; Marckmann, C.J.; Nyg{\aa}rd, J.; S{\o}renson, C.B.; Beuscher, F.; Forchel, A.; Michel, M. {\it Phys. Rev. B} {\bf 2000}, {\it 62}, 10950.

\bibitem{RejecNat06} Rejec, T.; Meir, Y. {\it Nature} {\bf 2006}, {\it 442}, 900-903.

\bibitem{LauxSurfSci88} Laux, S.E.; Frank, D.J.; Stern, F. {\it Surf. Sci.} {\bf 1988}, {\it 196}, 101-106.

\bibitem{HirosePRL03} Hirose, K.; Meir, Y.; Wingreen, N.S. {\it Phys. Rev. Lett.} {\bf 2003}, {\it 90}, 026804.

\bibitem{ButtikerPRB90} B\"{u}ttiker, M. {\it Phys. Rev. B} {\bf 1990}, {\it 41}, 7906-7909.

\bibitem{WeiszPRB89} Weisz, J.F.; Berggren, K.-F. {\it Phys. Rev. B} {\bf 1989}, {\it 40}, 1325-1327.

\bibitem{KardynalPRB97} Kardyna{\l}, B.; Barnes, C.H.W.; Linfield, E.H.; Ritchie, D.A.; Nicholls, J.T.; Brown, K.M.; Jones, G.A.C.; Pepper, M. {\it Phys. Rev. B} {\bf 1997}, {\it 55}, 1966-1969.

\bibitem{ThomasAPL95} Thomas, K.J.; Simmons, M.Y.; Nicholls, J.T.; Mace, D.R.; Pepper, M.; Ritchie, D.A. {\it Appl. Phys. Lett.} {\bf 1995}, {\it 67}, 109-111.

\bibitem{KristensenJAP98} Kristensen, A.; Bo Jensen, J.; Zaffalon, M.; S{\o}rensen, C.B.; Reimann, S.M.; Lindelof, P.E.; Michel, M.; Forchel, A. {\it J. Appl. Phys.} {\bf 1998}, {\it 83}, 607-609.

\bibitem{LiangPRB99} Liang, C.-T.; Simmons, M.Y.; Smith, C.G.; Kim, G.H.; Ritchie, D.A.; Pepper, M. {\it Phys. Rev. B} {\bf 1999}, {\it 60}, 10687-10690.

\bibitem{IqbalArxiv12} Iqbal, M.J.; de Jong, J.P.; Reuter, D.; Wieck, A.D.; van der Wal, C.H. arXiv:1207.1331.

\bibitem{NixonPRB91} Nixon, J.A.; Davies, J.H.; Baranger, H.U. {\it Phys. Rev. B} {\bf 1991}, {\it 43}, 12638-12641.

\bibitem{AkisJPCM08} Akis, R.; Ferry, D.K. {\it J. Phys.: Condens. Matter} {\bf 2008}, {\it 20}, 164201.

\bibitem{BerggrenJPCM08} Berggren, K.-F.; Yakimenko, I.I. {\it J. Phys.: Condens. Matter} {\bf 2008}, {\it 20}, 164203.

\bibitem{KowalskiPRB94} Kowalski, B.; Omling, P.; Meyer, B.K.; Hofmann, D.M.; Wetzel, C.; H\"{a}rle, V.; Scholz, F.; Sobkowicz, P. {\it Phys. Rev. B} {\bf 1994}, {\it 49}, 14786-14789.

\bibitem{RokhinsonPRL04} Rokhinson, L.P.; Larkina, V.; Lyanda-Geller, Y.B.; Pfeiffer, L.N.; West, K.W. {\it Phys. Rev. Lett.} {\bf 2004}, {\it 93}, 146601.

\bibitem{GlazmanJETP88} Glazman, L.I.; Lesovik, G.B.; Khmel'nitskii, D.E.; Shekter, R.I. {\it JETP Lett.} {\bf 1988}, {\it 48}, 238-241.

\bibitem{SzaferPRL89} Szafer, A.; Stone, A.D. {\it Phys. Rev. Lett.} {\bf 1989}, {\it 62}, 300-303.

\bibitem{LindelofJPCM08} Lindelof, P.E.; Aagesen, M. {\it J. Phys.: Condens. Matter} {\bf 2008}, {\it 20}, 164207.

\bibitem{WuPRB12} Wu, P.M.; Li, P.; Zhang, H.; Chang, A.M.; {\it Phys. Rev. B} {\bf 2012}, {\it 85}, 085305.

\bibitem{CrookSci06} Crook, R.; Prance, J.; Thomas, K.J.; Chorley, S.J.; Farrer, I.; Ritchie, D.A.; Pepper, M.; Smith, C.G. {\it Science} {\bf 2006}, {\it 312}, 1359-1362.

\bibitem{ThomasPM98} Thomas, K.; Nicholls, J.T.; Simmons, M.Y.; Pepper, M.; Mace, D.R.; Ritchie, D.A. {\it Phil. Mag. B} {\bf 1998}, {\it 77}, 1213-1218.

\bibitem{KirczenowPRB89} Kirczenow, G. {\it Phys. Rev. B} {\bf 1989}, {\it 39}, 10452-10455.

\bibitem{KomijaniEPL10} Komijani, Y.; Csontos, M.; Shorubalko, I.; Ihn, T.; Ensslin, K.; Meir, Y., Reuter, D., Wieck, A.D. {\it Europhys. Lett.} {\bf 2010}, {\it 91}, 67010.

\bibitem{KlochanPRL11} Klochan, O.; Micolich, A.P.; Hamilton, A.R.; Trunov, K.; Reuter, D.; Wieck, A.D. {\it Phys. Rev. Lett.} {\bf 2011}, {\it 107}, 076805.

\bibitem{SfigakisPRL08} Sfigakis, F.; Ford, C.J.B; Pepper, M.; Kataoka, M.; Ritchie, D.A.; Simmons, M.Y. {\it Phys. Rev. Lett.} {\bf 2008}, {\it 100}, 026807.

\bibitem{ChenPRB09} Chen, T.-M.; Graham, A.C.; Pepper, M.; Farrer, I.; Ritchie, D.A. {\it Phys. Rev. B} {\bf 2009}, {\it 79}, 153303.

\bibitem{SarkozyPRB09} Sarkozy, S.; Sfigakis, F.; Das Gupta, K.; Farrer, I.; Ritchie, D.A.; Jones, G.A.C.; Pepper, M.; {\it Phys. Rev. B} {\bf 2009}, {\it 79}, 161307.

\bibitem{YoonPRL07} Yoon, Y.; Mourokh, L.; Morimoto, T.; Aoki, N.; Ochiai, Y.; Reno, J.L.; Bird, J.P. {\it Phys. Rev. Lett.} {\bf 2007}, {\it 99}, 136805.

\bibitem{YoonPRB09} Yoon, Y.; Kang, M.-G., Morimoto, T.; Mourokh, L.; Aoki, N.; Reno, J.L.; Bird, J.P.; Ochiai, Y. {\it Phys. Rev. B} {\bf 2009}, {\it 79}, 121304.

\bibitem{YoonPRX12} Yoon, Y.; Kang, M.-G.; Morimoto, T.; Kida, M.; Aoki, N.; Reno, J.L.; Ochiai, Y.; Mourokh, L.; Fransson, J.; Bird, J.P. {\it Phys. Rev. X} {\bf 2012}, {\it 2}, 021003.

\bibitem{KouwenhovenRPP01} Kouwenhoven, L.P.; Austing, D.G.; Tarucha, S. {\it Rep. Prog. Phys.} {\bf 2001}, {\it 64}, 701-736.

\bibitem{HansonRMP07} Hanson, R.; Kouwenhoven, L.P.; Petta, J.R.; Tarucha, S.; Vandersypen, L.M.K. {\it Rev. Mod. Phys.} {\bf 2007}, {\it 79}, 1217-1265.

\bibitem{MeirPRL02} Meir, Y.; Hirose, K.; Wingreen, N.S. {\it Phys. Rev. Lett.} {\bf 2002}, {\it 89}, 196802.

\bibitem{GoldhaberNat98} Goldhaber-Gordon, D.; Shtrikman, H.; Mahalu, D.; Abusch-Magder, D.; Meirav, U.; Kastner, M.A.; {\it Nature} {\bf 1998}, {\it 391}, 156.

\bibitem{MeirPC} Meir, Y.; Private communication.

\end{thebibliography}
\end{document}